\documentclass{aa} 
\usepackage{txfonts}
\usepackage{graphicx}
\usepackage{adjustbox}
\usepackage[amssymb]{SIunits}
\usepackage{subfigure}
\usepackage{amsmath}
\usepackage[utf8]{inputenc}
\usepackage{natbib}
\usepackage{mathrsfs}
\bibpunct{(}{)}{;}{a}{}{,}

\begin{document}

\title{UV slope of z$\sim$3 bright ($L>L^{*}$) Lyman-break galaxies in the COSMOS field}
\author{S. Pilo\inst{1}
\and M. Castellano\inst{1}
\and A. Fontana\inst{1}
\and A. Grazian\inst{1}
\and K. Boutsia\inst{2}
\and L. Pentericci\inst{1}
\and E.Giallongo\inst{1}
\and E. Merlin\inst{1}
\and D. Paris\inst{1}
\and P. Santini\inst{1}}

\institute{INAF-Osservatorio Astronomico di Roma, Via Frascati 33, 00078 Monte Porzio Catone (RM), Italy \email{stefano.pilo@inaf.it}
\and Carnegie Observatories, Las Campanas Observatory, Casilla 601, La Serena, Chile }

\date{Received  / Accepted}

\titlerunning{z$\sim$3 LBGs UV slope in COSMOS}
\authorrunning{S.Pilo et al.}

\abstract
{The analysis of the UV slope $\beta$ of Lyman-break galaxies (LBG) at different luminosities and redshifts is fundamental for understanding their physical properties, and in particular, their dust extinction.} 
{We analyse a unique sample of 517 bright ($L>L^{*}$) LBGs at redshift z$\sim$3 in order to characterise the distribution of their UV slopes $\beta$ and infer their dust extinction under standard assumptions.}
{We exploited multi-band observations over 750 arcmin$^2$ of the COSMOS field that were acquired with three different ground-based facilities: the Large Binocular Camera (LBC) on the Large Binocular Telescope (LBT), the Suprime-Cam on the SUBARU telescope, and the VIRCAM on the VISTA telescope (ULTRAVISTA DR2). Our multi-band photometric catalogue is based on a new method that is designed to maximise the signal-to-noise ratio in the estimate of accurate galaxy colours from images with different point spread functions (PSF). We adopted an improved selection criterion based on deep Y-band data to isolate a sample of galaxies at $z\sim 3 to$ minimise selection biases. We measured the UV slopes ($\beta$) of the objects in our sample and then recovered the intrinsic probability density function of $\beta$ values (PDF($\beta$)), taking into account the effect of observational uncertainties through detailed simulations.}
{The galaxies in our sample are characterised by mildly red UV slopes with $<\beta>\simeq -1.70$ throughout the enitre luminosity range that is probed by our data ($-24\lesssim M_{1600}\lesssim -21$). The resulting dust-corrected star formation rate density (SFRD) is $log(SFRD)\simeq-1.6 M_{\odot}/yr/Mpc^{3}$, corresponding to a contribution of about 25\% to the total SFRD at z$\sim$3 under standard assumptions.}
{Ultra-bright LBGs at $z \sim 3$ match the known trends, with UV slopes being redder at decreasing redshifts, and brighter galaxies being more highly dust extinct and more frequently star-forming than fainter galaxies.}

\maketitle          

\section{Introduction}
\label{section1}
 
The introduction of the Lyman break technique more than 25 years ago (e.g. \citealp{sh1,sh2}) has enabled the selection of UV bright star-forming galaxies at z$\gtrsim$3, thus  opening a window on the earliest phases of galaxy formation at high redshift. Since then, the  Lyman break method has been exploited to select objects at increasingly fainter magnitudes and higher redshifts, which extended our knowledge on the statistical properties of distant galaxy populations, such as the UV luminosity function (LF) \citep[e.g.][]{Bouwens2015,Finkelstein2015,Ono2018,Oesch2018}, the size distribution, and the size-luminosity relation \citep[e.g.][]{Grazian2012,Kawamata2015,Curtis-Lake2016}. 

Constraining the physical properties of Lyman-break galaxies (LBG) is fundamental for converting these global properties of the galaxy populations into global physical properties: most importantly, for inferring the total star-formation rate density (SFRD) from the  UV LF. In practice, estimating the total SFRD requires knowing the conversion factor between the UV luminosity and the star formation rate (SFR), and of the amount of extincted UV radiation. The conversion between L(UV) and SFR has been routinely fixed on the basis of stellar population synthesis models \citep[e.g.][]{Madau1998}, while the slope $\beta$ of the power-law UV continuum has acquired increasing importance for estimating dust extinction \citep[e.g.][]{calz1994,meu1999,calz2000}. The UV slope is also affected by metallicity, age, star formation history, and stellar initial mass function, although dust extinction likely remains the dominant reddening factor \citep[e.g.][]{Wilkins2013,Castellano2014}. While the conversion between observed luminosity and SFR is further complicated by uncertainties on the shape of the attenuation law of high-redshift galaxies \citep[e.g.][]{Capak2015,reddy2018,McLure2018,Koprowski2018}, accurate measurements of the UV slope of large samples of LBGs, and the determination of relations between UV slope, luminosity, and redshift, remain a fundamental ingredient for constraining the evolution of galaxies and the SFRD in the first $\sim2$ Gyr after the Big Bang \citep{Finkelstein2012,Bouwens2014}. Unfortunately, discrepancies among different works remain on the  $\beta - L(UV)$ and  $\beta - redshift$ relations, which might be explained by differences in selection criteria and by a poor evaluation of the impact of observational effects and selection biases (e.g. \citealp{bou2009,Bouwens2012,cast2012,Dunlop2013}). 

In this paper we analyse the UV slope of a large sample of bright ($L\gtrsim L^*$) galaxies at z$\sim$3, that is, at the peak of the SFRD. 
Compared to previous analyses of UV slopes at z$\sim$3 \citep{bou2009,fink2012,hathi2013,Kurczynski2014,Pannella2015}, we exclusively focus on a large sample of galaxies in the high-luminosity range. In addition, we introduce new colour-colour criteria that combine optical and IR data and are aimed at an accurate and efficient selection of moderately extincted LBGs. The determination of the UV slope distribution is carried out using the technique presented in \citet{cast2012} (C12 hereafter), which is meant to recover the intrinsic probability distribution function PDF($\beta$) taking into full account observational and selection effects. Particular care has also been taken in the construction of a multi-band photometric catalogue that maximises the accuracy of colours and UV slope measurements from imaging data at different resolution.

The structure of the paper is as follows: in Sect.~\ref{section2} we present the dataset exploited in this paper; in Sect.~\ref{section3} we describe the selection criterion we use for the extraction of the galaxy sample and assess its completeness and efficiency, and Sects. ~\ref{section4} and ~\ref{section5} present the results on PDF($\beta$) and SFRD measurements, respectively. Summary and conclusions (Sect. ~\ref{section7}) follow. A detailed description of the new photometric method we used to extract the source catalogues is given in the appendix.  Throughout the whole paper, observed and rest frame magnitudes are in the AB system, and we adopt the $\Lambda$-CDM concordance model ($H_{0}=70\ km\ s^{-1}\ Mpc^{-1}$, $\Omega_{M}=0.3$ and $\Omega_{\Lambda}=0.7$).

 \section{Multi-band observations}
 \label{section2}
We exploit imaging data in nine different bands over a common area of $\sim 750\ arcmin^2$ of the COSMOS field \citep{scov2007}. The dataset covers a spectral range from U band ($\lambda_{central}$=355nm) to K band ($\lambda_{central}$=2150nm), and it has been acquired by three different ground-based instruments. 
The U$_{special}$ filter, the G-Sloan, and the R-Sloan{$_{blue}$} images were obtained with the Large Binocular Camera \citep{gial2008} at the LBT on Mount Graham in Arizona, and are described in \citet{konst2014} \citep[see also][]{Grazian2016,Grazian2017}. The I- and Z-band filter images \citep{Taniguchi2007,capak2007} have been acquired by the Suprime-Cam \citep{miya2002} mounted on the Subaru Telescope at Maunakea. The J-, Y-, H-, and K-band images are from the second data release of the UltraVISTA survey \citep[][]{mccra2012}. The U, G, and R data from LBC are significantly deeper than available Subaru data at similar wavelengths. In practice, we restrict our analysis to the portion of the full COSMOS area where deep LBC observations enable an accurate sampling of the Lyman break at z$\sim$3. 
The main properties of our dataset are summarised in Table~\ref{tab1}.

We used the LBC R band as detection image and extracted a multi-band photometric catalogue from all bands according to a new technique, which is described in detail in the Appendix. We used SExtractor \citep{bertarn1996} to measure the total (Kron) flux in the detection band, while total fluxes in the other bands were measured by scaling the total flux of the detection band according to the relevant colour terms. The novelty of our approach lies in the criteria we adopted to measure colours. 

Briefly, we estimated the colour between the measure and detection images using optimally chosen apertures, scaled on the basis of the relevant PSF-FWHM in order to always recover  the same fraction of flux in all bands for each source. This procedure allows the photometry to be extracted without resorting to PSF-matching techniques, thus preserving the original resolution in each image and avoiding any degradation of the photometric information. As shown in the appendix, the adopted technique enables colour (and UV slope) measurements at a signal-to-noise ratio (S/N) that is higher than can be achieved with PSF-matching techniques. The final catalogue includes 45831 objects down to R$\sim$28 (being R=26.4 at S/N=10, see Table \ref{tab1}).

\begin{table*}       
\centering      
\begin{adjustbox}{max width=\textwidth}             
\begin{tabular}{ccccccc} 
\hline    
\hline                 
Filter & $\lambda_{central}($nm$)$ & Exp. Time (s) & FWHM ($arcsec$)  & AB Mag.limit (S/N=10) & Instrument & Pixel scale ($\frac{arcsec}{pixel}$) \\    
\hline                        
$U_{special}^a$ & 355 & 28700 & 0.94      &  26.6 (26.8) & LBC (LBT) & 0.225 \\     
$G_{sloan}^a$  &  475 &  12200 & 1.12     & 27.1 & LBC (LBT) & 0.225 \\
$R_{sloan}^a$  &  622 &  12000 &  0.97   & 26.4 & LBC (LBT) & 0.225 \\ 
$I_{sloan}^b$  & 764 & 27000 & 0.97   & 25.8 (26.2) & Suprime-Cam (Subaru) & 0.150 \\ 
$Z_{sloan}^b$  & 903 &  38880 & 1.15    &  24.9 & Suprime-Cam (Subaru) & 0.150 \\ 
$Y^c$  & 1020 &  42360 & 0.85       &  25.1 (26.0) & Vir-Cam (VISTA) & 0.300 \\ 
$J^c$  & 1250 &  49720 & 0.81       & 24.9 (25.5) & Vir-Cam (VISTA) & 0.300 \\
$H^c$ & 1650 &  42520 & 0.80       & 24.6 (25.0) & Vir-Cam (VISTA) & 0.300 \\
$K^c$ & 2150 &  39400 & 0.80       & 24.1 (25.1) & Vir-Cam (VISTA) & 0.300 \\
\hline      
\end{tabular}
\end{adjustbox}
\caption{Imaging dataset. The FWHM of the PSF has been measured on a stacking of a selected sample of bright non-saturated stars. Limiting magnitudes have been measured in diameter apertures of $2\ arcsec$ . Because some images have variable depth across the field, we report in parentheses the limiting magnitude of the deepest portions in each band. References: a) \citet{konst2014}, b) \citet{Taniguchi2007}, and c) \citet{mccra2012}.}
\label{tab1}
\end{table*}

\section{z$\sim$3 sample}
\label{section3}
\subsection{New selection criterion for sampling U-dropout LBGs}
We applied a tailored version of the Lyman-break technique \citep{steid2003} to select LBGs at z$\sim$3. The standard colour selection is based on a (U-G) versus (G-R) diagram (UGR criterion hereafter, e.g. \citealp{sh1,sh2,Giavalisco2002,konst2014}), where the U and G bands sample the $912\ \angstrom$ break, and the (G-R) colour samples the UV continuum of z$\sim$3 star-forming galaxies. We used a different approach that takes advantage of the deep Y band instead of the R band. While selection criteria based on optical data alone, such as the UGR one, provide advantages in terms of observing efficiency and limiting depth, the availability of new, efficient IR detectors has enabled adopting IR colours in the selection of distant sources. It has indeed been shown (e.g. \citealp{beckw2006,cast2012}, C12 hereafter) that exploiting a red filter improves the selection efficiency because it avoides contamination in the sample from dusty star-forming and passive galaxies at lower redshift. 
 
Colour criteria were defined on the basis of models from Charlot and Bruzual 2007 (\citealp{bc2007a,bc2007b}, hereafter CB07) to efficiently separate LBGs from low-redshift interlopers:

\begin{gather*} 
U-G > 1.0\ \wedge\ U-G > (G-Y) + 1.2.
\end{gather*}

\begin{figure*}
\centering
 \includegraphics[width=6.2cm]{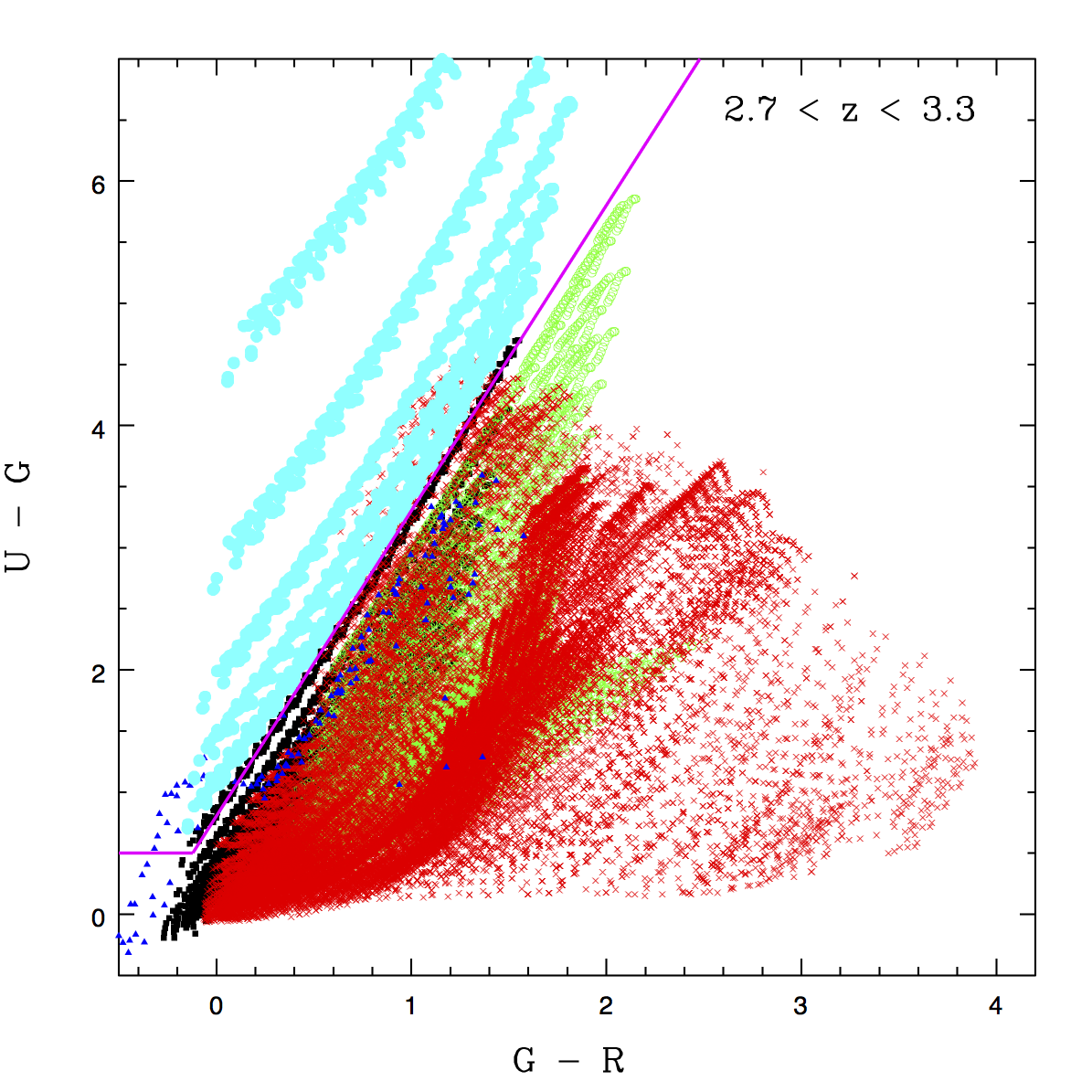}
 \includegraphics[width=6.2cm]{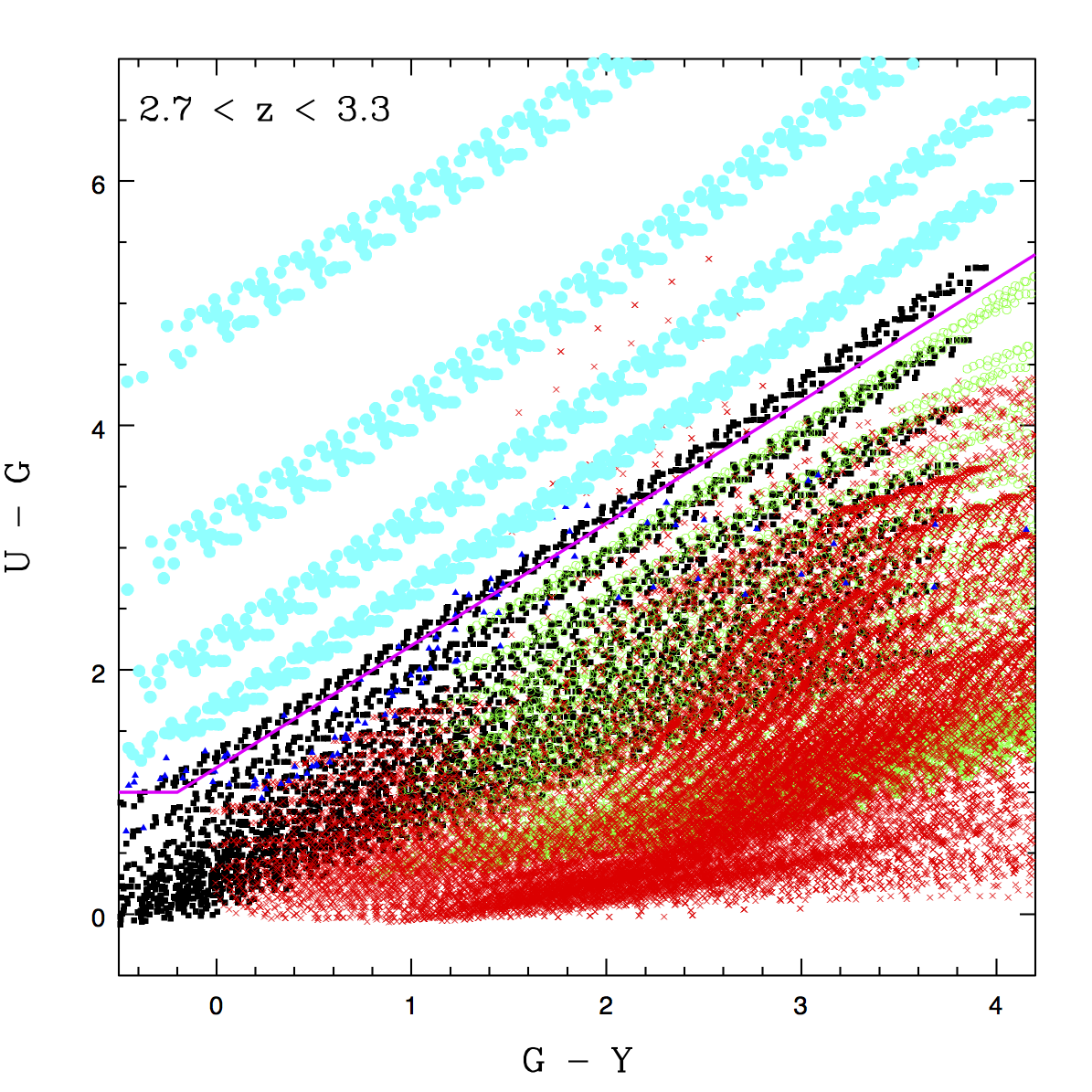}
\hspace{2mm}
\caption{Comparison between the $UGR$ (left panel) and $UGY$ criteria (right panel). Different colours and shapes mark the position of galaxy models. Cyan points mark the LBGs in the redshift range $2.7<z<3.3$ with $Z/Z_{\odot}=0.2-1.0$, $age=0.01,0.1,0.3,0.5,1,2\ Gyr$, constant SFR, at an increasing extinction $E(B-V)=0-1.0$ at 0.05 intervals; black filled squares indicate LBGs out of the desired redshift range; red crosses indicate red passively evolving galaxies at lower redshifts, with $Z/Z_{\odot}=0.2-1.0$, $age=1-13\ Gyr$, $\tau=0.1,0.3,0.6$ at an increasing extinction $E(B-V)=0-0.2$ at 0.05 intervals; green open circles mark low-redshift dusty star-forming galaxies with $Z/Z_{\odot}=0.2-1.0$, $age=0.01-1\ Gyr$, constant SFH at an increasing extinction $E(B-V)=0.5-1.5$ at 0.1 intervals; stars from the \cite{Pickles1985} library are shown as blue filled triangles. The magenta continuous lines indicate the LBG selection windows.}
\label{fig5}
\end{figure*}

These cuts were set after they were optimised on the basis of the analysis of spectroscopic redshifts in order to include as many sources as possible in the desired redshift range while avoiding the contamination of interlopers.

In Fig. \ref{fig5} we compare the standard UGR selection criterion to the adopted UGY criterion, showing the positions of star-forming and passive galaxies with different ages, extinctions, metallicities, and with constant and also exponentially declining star formation histories (SFH). The UGY criterion clearly provides a clean selection of the LBGs in the target redshift range, regardless of their physical properties, and it enables the definition of conservative cuts to avoid contaminants. Low-redshift contaminants, especially red passive galaxies, are closer to moderate E(B-V) LBGs in the UGR diagram than in the UGY diagram.  A re-definition of the UGR selection window to make it more robust against photometric scatter appears unfeasible.

To avoid including objects with unreliable $\beta$ values due to noisy photometry, we applied cuts at $R\leq24.8$ and $Y \leq 25.3,$ corresponding to an $R-Y = -0.5$ colour. This colour cut corresponds to $\beta\geq -3.0$, thus including all slope values that are predicted by standard stellar libraries. In practice, these criteria isolate objects with $S/N(R)\gtrsim 20$ and with $S/N(Y)\gtrsim 5-10$ in the shallower and in the deeper Y-band stripes, respectively. The observed R magnitude cut corresponds to a rest-frame cut at $M_{1600}\simeq -20.8$.

\begin{figure}
\centering
 \includegraphics[width=8.0cm]{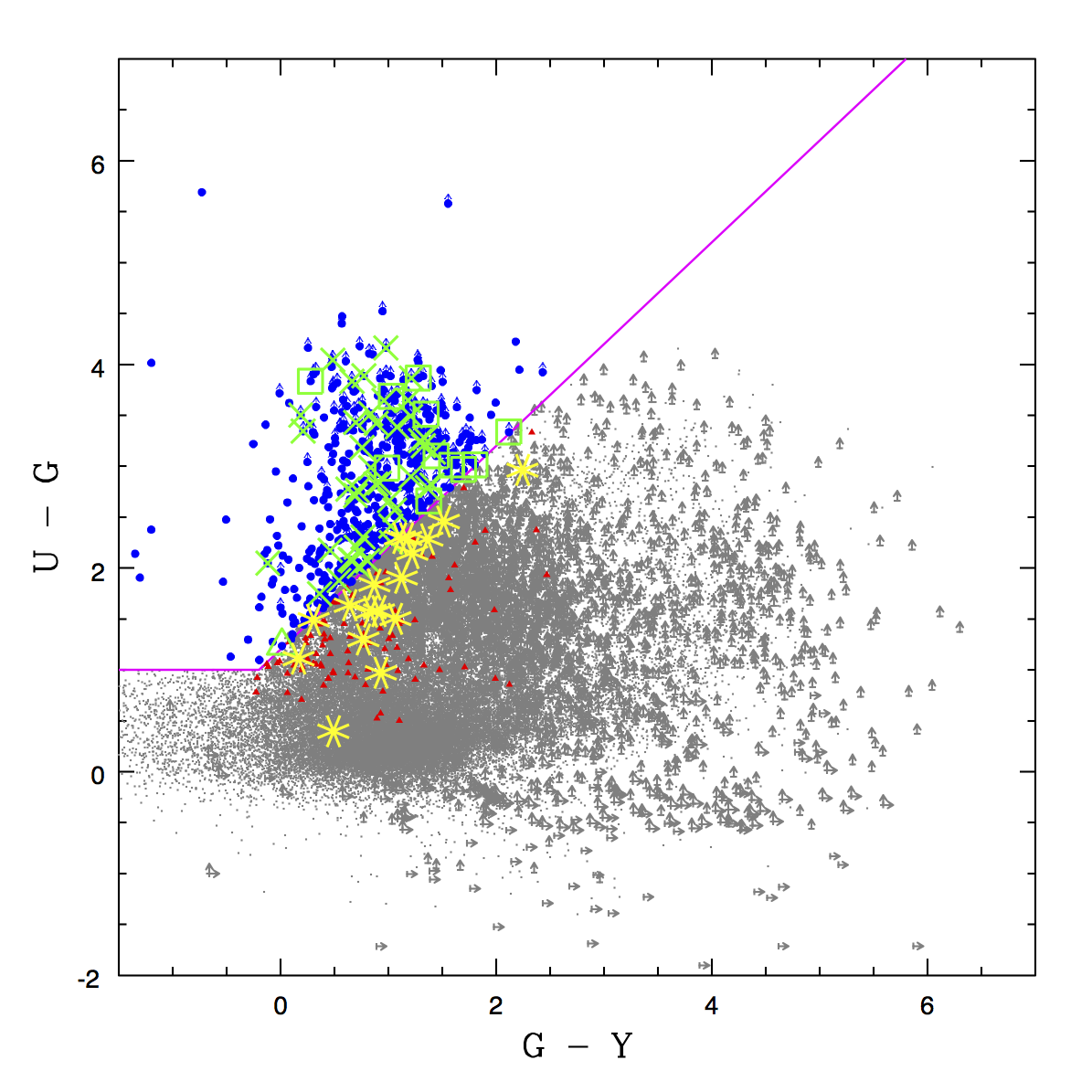}
\caption{$UGY$ selected objects (blue points) along with objects selected with the $UGR$ criterion that are excluded by the UGY criterion (red triangles). Upwards-pointing and right-pointing arrows mark upper limits in the U and G band, respectively.  
Objects with spectroscopic redshift in the range $2.7<z<3.3$ are marked as green crosses. Open triangles and open squares indicate sources at $z<2.7$ and $z>3.3$, respectively. Objects in the expected redshift range that are excluded by our selection are indicated in yellow.}
\label{fig6}
\end{figure}

This sample has been further refined by excluding
\begin{itemize}
\item sources that were identified as stars by visual inspection after checking  all objects with SExtractor $CLASS\_STAR>0.95$,
\item sources that were found to be X-ray emitters after cross-correlating our catalogue with \cite{mccracken2012} and \cite{capak2007},
\item objects that were not properly deblended, which have been detected as a unique source but at visual inspection were clearly found to be the blending of two distinct sources in at least one of the available bands,  
\item objects falling on artefacts, bad pixels, or at the very border of the image in at least one of the available bands, and 
\item objects falling into the halo of a nearby very bright saturated star in at least one of the available bands. 
\end{itemize}

The final sample consists of 517 objects at $21.8\lesssim R\lesssim 24.8$. The criteria described above ensure that our galaxy sample is not contaminated by stars, bright AGN (except any X-ray undetected AGN), or by spurious and problematic objects.

In figure \ref{fig6} we show the relevant colour-colour diagrams and the differences between the criteria adopted in the present paper and the usual UGR selection. Our selected candidates are indicated by blue points, while the red points mark the position in the diagram of objects that would have been included by a standard UGR colour-selection, but are excluded from our sample.

We used the \texttt{zphot.exe} code with the well-tested procedure described in \cite{font2000} and \cite{graz2006} (see also \citealp{dahl2013} and \citealp{sant2015}) to compute photometric redshifts for our objects. We find an average $<z>\simeq3.2$ for our LBG sample, in agreement with the expected redshift selection range.

\subsection{Spectroscopic redshift validation of the sample}\label{specval}
We cross-correlated our catalogue with the VUDS spectroscopic catalogue (\citealp{LeFevre2014}) and the DEIMOS 10K Spectroscopic Survey Catalog of the COSMOS Field (\citealp{Hasi2018}) to measure completeness in the target redshift range, and potential contamination by lower and higher redshift galaxy interlopers.
We performed the cross-correlation in a $0.97\ arcsec$ diameter corresponding to the R-band PSF-FWHM. We chose to take into account only sources with flags 3 and 4 (reliability $\geq 95$\%). We find that 54 sources in our UGY colour-selected sample ($\sim 11\%$) have a spectroscopic redshift: 42 objects ($\sim80\%$) fall in the reference redshift range $2.7<z<3.3$ (the same as  was used for evaluating the colour cuts from the libraries), 1 object ($2\%$) is found to be at slightly lower redshift ($z=2.68$), and 11 objects ($18\%$) lie at $z=3.32-3.77$. No contamination from low-redshift interlopers is found, consistently with the expectation that the UGY criterion efficiently excludes red and dusty low-redshift galaxies. Seventeen sources at $2.7<z<3.3$ fall within our magnitude cuts but are not selected by our colour criteria: we find that all of them lie very close to the selection region in the colour-colour diagram and are excluded due to photometric scatter. 

This scatter is inevitable when a pure colour selection criterion is applied because it stems from a combination of photometric noise and spectral variations of the sources. In Sect.~\ref{statisticalanalysis} we describe how these observational uncertainties are fully taken into account in the estimate of UV slope distributions. In any case, we have verified that a modification of the colour cut in order to include these objects would worsen the contamination from low-redshift interlopers in the sample. 
We also  verified that our UGY colour selection minimises this effect compared to the standard UGR. With this purpose, we repeated the same test using the standard UGR colour criterion. We find a lower selection efficiency with only 33 sources at $2.7<z<3.3$ included (together with 6 interlopers), and 26 sources excluded. We find that the UGR selection includes interlopers at a lower redshift ($z\sim2.4-2.5$) than the UGY criterion. We conclude that the UGY selection is both more efficient and more robust than the standard UGR selection, at least at the depth of our dataset. Figure \ref{fig6} shows the position of spectroscopic objects in our sample in the UGY colour-colour plane.

\subsection{Selection completeness}\label{complet}
\begin{figure}
\centering
 \includegraphics[width=8.0cm]{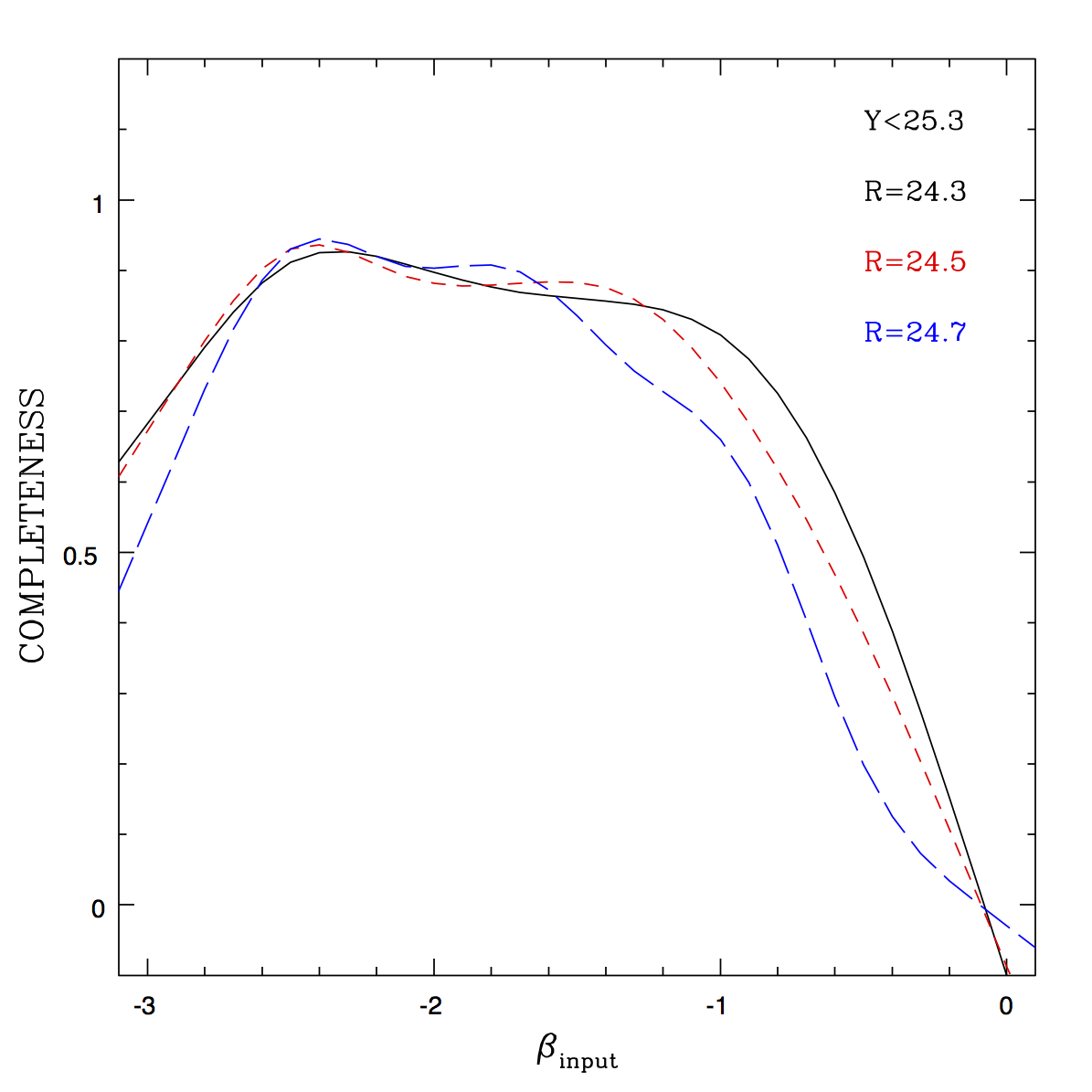}
\caption{Detection completeness as a function of $\beta_{input}$ as found from our Monte Carlo simulations for sources with Y$<$25.3 in magnitude bins centred at mag $R = 24.3, 24.5, 24.7$ (black continuous line, red short-dashed line, and blue long-dashed line, respectively).}
\label{fig7}
\end{figure}

We used Monte Carlo simulations to check the completeness of our sample as a function of $\beta$, which is measured from the I, Z, Y magnitudes as described in Sect.~\ref{section4}. We used CB07 libraries to produce a set of simulated galaxies in the redshift range $2.5<z<3.5$. We considered galaxy models with constant SFHs and within the following range of physical parameters: $ Z/Z_{\odot} = 0.02, 0.2, 1.0 $ , $ 0.01\leq E(B-V)\leq 1.0 $, $ 0.01\text{}\leq Age\leq 1 Gyr $. We assumed a \cite{salp1955} initial mass function (IMF) and a \cite{calz2000} attenuation law, while the transmission of the intergalactic medium (IGM) was treated according to \cite{fan2006}. We generated a mock catalogue of $ 5.6 \times 10^5 $  objects by perturbing magnitudes in all the bands to match the relevant depths in the observed bands. The simulated catalogues were then treated and analysed like the real catalogues in order to evaluate the observational effects on galaxies of different magnitudes and spectral slopes. 
We evaluated the detection completeness of galaxies with a given input spectral slope $\beta_{input}$ in different magnitude bins. 
Figure \ref{fig7} shows that the applied cuts yield a completeness of $\sim$90\% at $-2.8 \lesssim \beta \lesssim -1.0$ (corresponding to $E(B-V)\lesssim 0.25$, following \citealt{calz2000}) up to $R\simeq 24.5$. The completeness at $R\simeq 24.7$ decreases slightly more rapidly at red UV slopes but is still $\sim$70\% for $\beta\sim -1$. The 50\% completeness limit lies at  $\beta\sim-0.4 - -0.8$ ($E(B-V)\lesssim 0.30 - 0.39$) depending on the observed R-band magnitude. A similar test on a UGR-selected sample shows lower completeness levels for red objects, namely 50\% completeness limits at $\beta\sim-0.7- -1.0$. In conclusion, we find that our selection criteria enable the analysis of a large range of physically meaningful values of $\beta$ over the chosen range of magnitudes, and  they improve the selection of red sources with respect to the standard UGR criteria.

\section{UV slopes of z$\sim$3 bright LBGs}
\label{section4}

\subsection{Measurement of the UV slopes}
We assumed the UV spectrum of the LBGs to be a power law $F_{\lambda}=\lambda^{\beta}$ and measured $\beta$ by performing a linear fit over the observed I, Z, Y magnitudes (spanning the rest-frame wavelength range $\lambda \simeq 1750-2750\ \angstrom$ at z$\sim$3): $M_{i}=-2.5(\beta+2.0)log(\lambda_{i})+c$, where $M_{i}$ is the magnitude measured in the \textit{i}th filter whose effective wavelength is $\lambda_{i}$. 
Uncertainties on the observed magnitudes were taken into account in deriving $\beta$ and its uncertainty, whose typical value is $\sigma_{\beta}\simeq 0.2$. We visually inspected a handful of objects (mostly with steep slopes) with larger errors (up to $\sigma_{\beta} \simeq 0.6-0.8$) to ensure that the larger uncertainty on $\beta$ is due to a lower S/N rather than to systematics in their flux estimates.

\begin{figure}
\centering
\includegraphics[width=8.0cm]{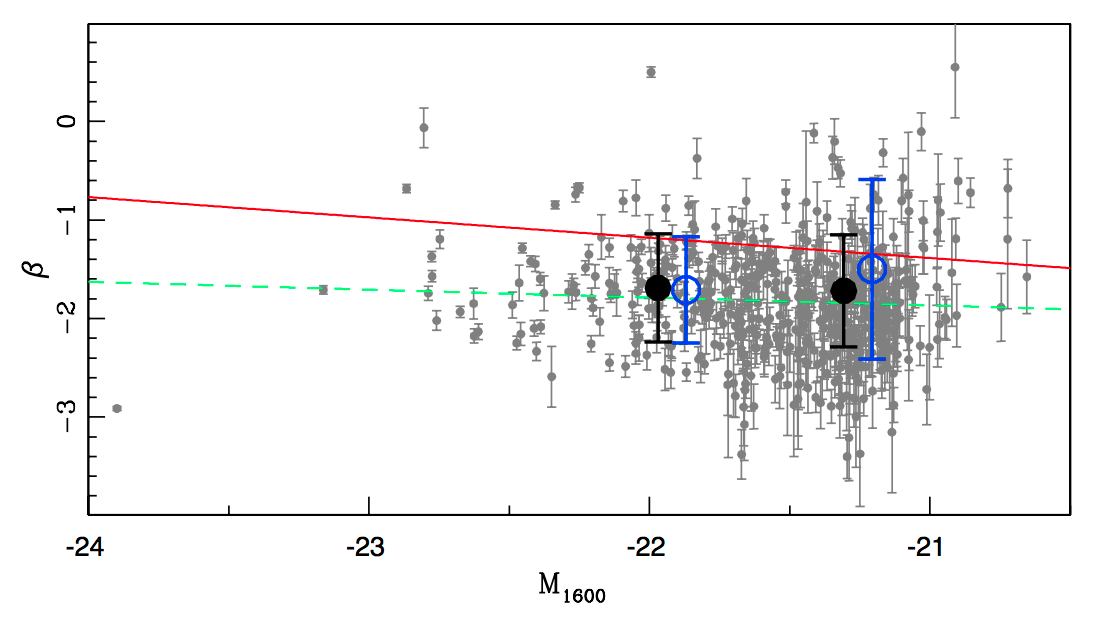}
\caption{UV slope as a function of $M_{1600}$. Black filled (blue open) circles show the best-fit values of the Gaussian (log-normal) distributions in each magnitude bin along with the relevant 1$\sigma$ scatter. The mean $\beta$ value is at $<\beta> = -1.82 $ . The red line is the $\beta$ - UV relation by \cite{bou2009}. The green dashed line marks the linear fit to our data.}
\label{fig12}
\end{figure}

The resulting observed relation between $\beta$ and UV rest-frame magnitude at $1600\ \angstrom$ is shown in Fig.~\ref{fig12}. A linear fit on the observed relation is shown as the green dashed line in Fig.~\ref{fig12} and suggests little variation of the UV slope with luminosity: $\beta=(0.003\pm0.005)\cdot M_{1600} -1.58 \pm 0.11$ (Spearman correlation coefficient $r_s=-0.09$).
However, as discussed in C12 and highlighted in Sect.~\ref{complet} of this paper, the observed magnitudes and colours affect the accuracy and completeness of UV slope estimates significantly. This implies that to measure the dependence of $\beta$ on UV magnitude, and in particular to recover the probability distribution function of $\beta$ at different luminosities, it is not sufficient to perform a straightforward fit to the observed distributions, but it is necessary to de-convolve such distributions from observational effects.

\subsection{Recovering the intrinsic UV slope distributions}\label{statisticalanalysis}

We followed the approach described in C12, which exploits Monte Carlo simulations to take into account all observational effects and any dependence of selection completeness on flux and colour in the estimate of the probability distribution function of the UV slopes, PDF($\beta$).

In practice, 1) we assumed that the probability distribution function PDF($\beta$) follows a given functional form, whose average ($<\beta>$) and standard deviation (rms($\beta$)) have to be estimated in different UV magnitude ranges. 2) For each ($<\beta>$, rms($\beta$)) pair in the parameter space of interest, we extracted a large number of galaxy templates at $2.5<z<4.0$ and randomly perturbed their fluxes, matching in each band the relevant relation between magnitude and S/N from our observed catalogue. 3) We applied our LBG selection criteria and measured the UV slope distribution of such a mock sample in the same way as for the real one. 4) We assessed the likelihood of each assumed pair of parameters by comparing the simulated and observed $\beta$ distributions using a maximum likelihood estimator $\mathscr{L}$ (e.g. \citealp{bou2008}; \citealp{cast2010a}). Finally, the best-fit pair of parameters ($<\beta>$, rms($\beta$)) was found by minimising $\Delta\chi ^2=-2.0\ln(\mathscr{L})$.

This procedure was separately applied to two different magnitude bins in order to obtain information on how the intrinsic PDF($\beta$) varies with UV luminosity: the first bin is at $-24.0<M_{UV}\leq -21.6$ (210 objects) and the second at $-21.6<M_{UV}\leq-21$ (288 objects). The choice of the bin widths is somewhat arbitrary, but ensures a suitable number of objects for the statistical analysis of each bin.

\begin{figure}
	\centering
	\includegraphics[width=8.0cm]{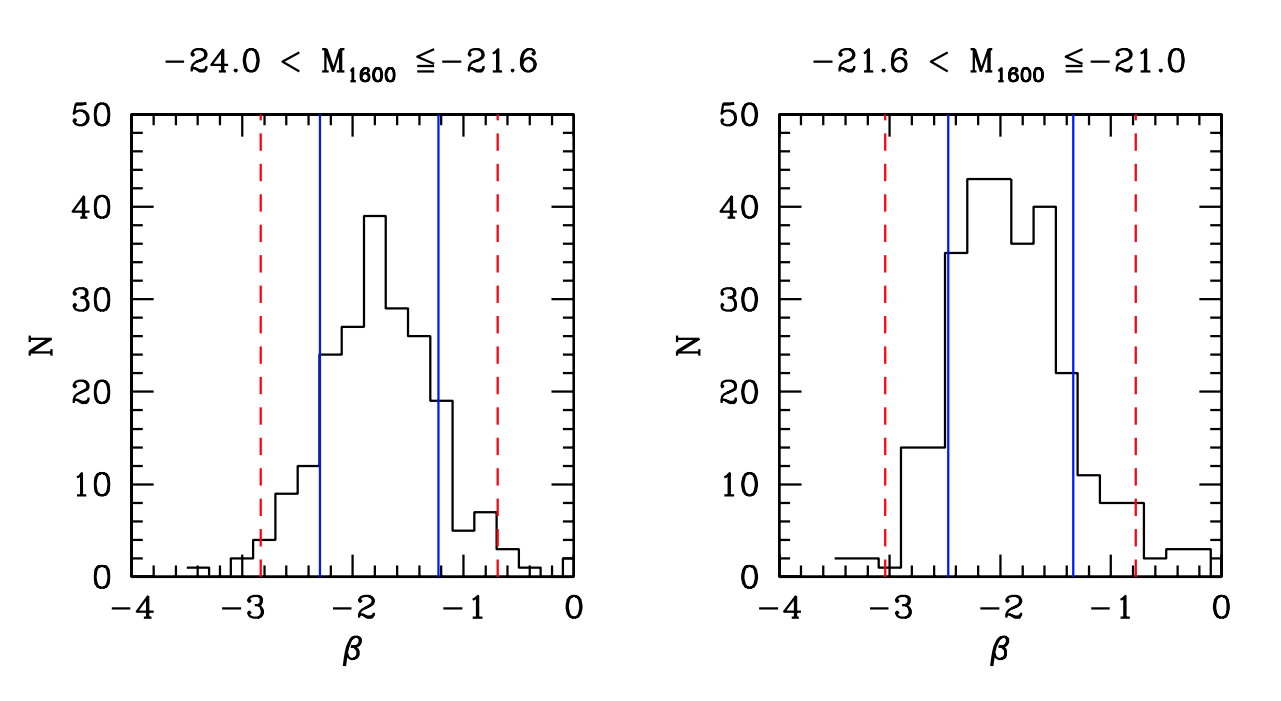}
        \caption{Observed UV slope distributions at $-24.0 < M_{1600} \leq -21.6 $ (left panel) and $ -21.6 < M_{1600} \leq -21.0 $ (right panel). The $1\sigma$ and $2\sigma$ dispersion ranges are included within the blue continuous and red dashed lines, respectively.}
        \label{fig10}
\end{figure}

We assumed as the shape of the PDF($\beta$) both a Gaussian (as has been suggested previously, e.g. \citealp{bou2009}) and a log-normal distribution. As was noted by C12, the choice of a log-normal PDF($\beta$) as alternative to a Gaussian is suggested by the shape of the observed UV slopes. As an example, the distribution in our faintest bin (right panel in Fig.~\ref{fig10}) shows a tail of red objects, which suggests a possible asymmetric distribution with a bulk of low-extinction LBGs along with only few sources at higher E(B-V) values.

Since the Monte Carlo simulations described in step 2) are time consuming, we first constrained the parameter space of interest by testing a wide range of ($<\beta>$, rms($\beta$)) values on a coarse sampling of the ranges $<\beta>$=-2.2, -1.0 and rms ($\beta$)=0.1, 1.5. We then accurately constrained their best-fit and confidence range through the analysis of a 35x35 grid in the range $<\beta>$= -2.0, -1.4 for both the Gaussian and log-normal distributions, and rms ($\beta$)=0.4, 0.8 (for the Gaussian distribution) and rms($\beta$)=0.4,1.0 (log-normal). For each position in the grid and for each magnitude bin, we extracted 25000 objects from the CB07 library and perturbed magnitudes in all bands as described in 3). 
We note that the estimate of $\mathscr{L}$ in step 4) requires the observed and simulated UV slopes distributions to be binned: the choice of the bin widths was tailored to ensure that the extended but poorly populated tails in the observed distributions are efficiently sampled by the simulations. As shown in Fig.~\ref{fig10}, different bin widths were thus set depending on the 1, 2, 3 $\sigma$ ranges of the observed distribution so as to reduce numerical noise in our procedure.


\begin{figure*}
\centering
 \includegraphics[width=6.2cm]{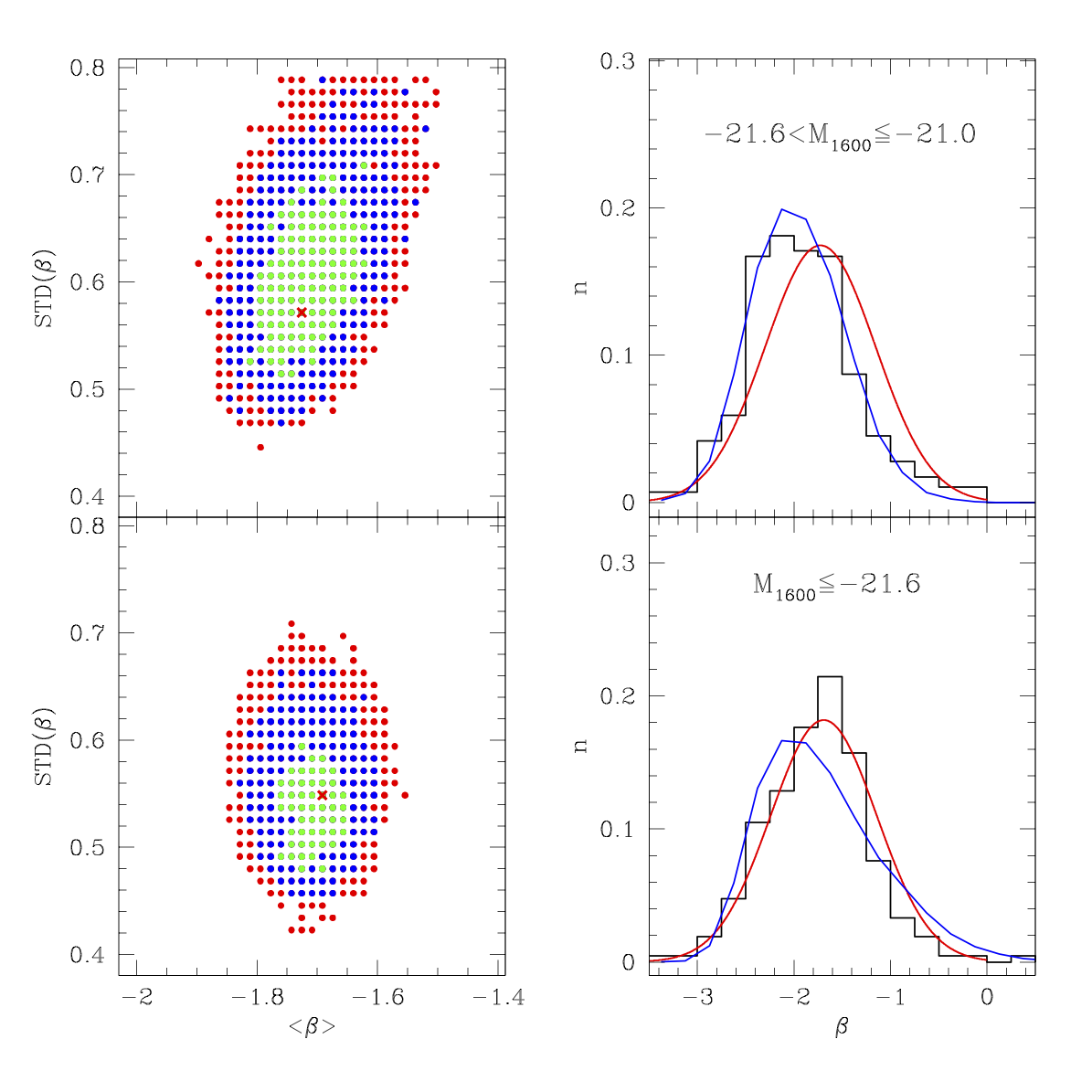}
 \includegraphics[width=6.2cm]{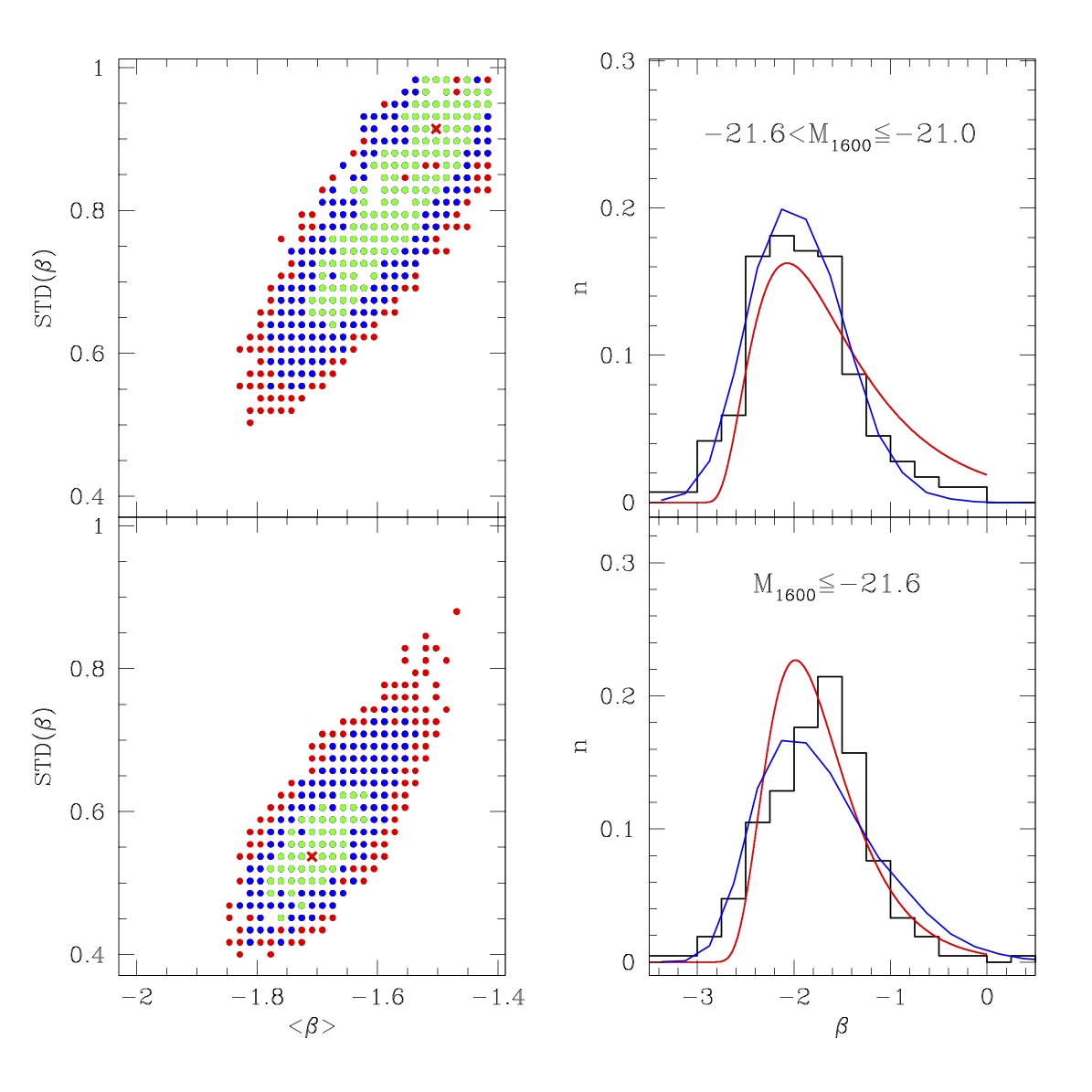}
\hspace{2mm}
\caption{Simulation results for the Gaussian (left panel) and log-normal (right panel) fit in the two different magnitude bins. In the leftmost column, we show maximum likelihood contours at 68\%, 95\%, and 99\% c.l. in green, blue, and red, respectively; the red cross indicates the position of the best fit parameters. The rightmost column shows the histogram of the observed data and the relevant best-fit distribution. The red continuous line is the intrinsic PDF($\beta$), which is transformed into the blue continuous line by observational effects. }
\label{fig11}
\end{figure*}

The best-fit parameters of the Gaussian and log-normal PDF($\beta$) in each magnitude bin are reported in Table \ref{tab2}. Confidence regions for the PDF($\beta$) are shown in Fig.~\ref{fig11}, together with observed and best-fit distributions. It is also highlighted how the latter (blue lines) are reshaped by observational effects (red): only a full modelling approach such as presented in C12 and adopted in this work allows us to recover the intrinsic slope distributions from observations. 

We find that $L>L^*$ objects at z$\sim$3 have typical, average UV slopes $\sim$-1.7. No significant variation with UV magnitude is found in the Gaussian case. The log-normal best-fit distributions point to a slightly redder average ($\sim$-1.5) in the fainter bin, although with large correlated uncertainties between the parameters. The absence of a strong correlation between UV slope and magnitude is in agreement with what was found by C12 at z$\sim$4, but it disagress with the previous analysis of z$\sim$3 objects by \cite{bou2009}.

While it is not possible to determine whether the best assumption is the Gaussian or the log-normal shapes, we note that the latter does not seem to accurately recover objects in the (poorly populated) left tail of the observed distribution in the brightest bin. This is also reflected in the noisier maximum likelihood contours.  

\begin{table*} 
\centering       
\begin{adjustbox}{max width=\textwidth}             
\begin{tabular}{ccc}
\multicolumn{3}{c}{Gaussian} \\
\hline    
\hline                 
UV magnitude &  <$\beta$> & rms($\beta$)  \\    
\hline          
$M_{1600}\leq -21.6$ & $-1.69_{-0.07}^{+0.03}$ & $0.55_{-0.07}^{+0.04}$  \\     
 $-21.6<M_{1600}\leq-21.0$ &  $-1.72_{-0.09}^{+0.10}$ & $0.57_{-0.06}^{+0.13}$ \\
\hline    
\end{tabular}
\end{adjustbox}

\vspace{2mm}

\begin{adjustbox}{max width=\textwidth}             
\begin{tabular}{cccc}
\multicolumn{4}{c}{Log-normal} \\
\hline    
\hline                 
UV magnitude& MAX &  <$\beta$> & rms($\beta$)  \\    
\hline                        
$M_{1600}\leq -21.6$& $-1.98_{-0.07}^{+0.05}$ & $-1.71_{-0.08}^{+0.06}$ & $0.54_{-0.07}^{+0.08}$  \\     
 $-21.6<M_{1600}\leq-21.0$ & $-2.07_{-0.09}^{+0.18}$ &  $-1.50_{-0.09}^{+0.08}$ & $0.91_{-0.13}^{+0.07}$ \\
\hline    
\end{tabular}
\end{adjustbox}
\caption{UV slope distribution: best-fit parameters and 68\% c.l. uncertainties.}
\label{tab2}
\end{table*}

\subsection{Comparison with previous results}
Investigations on the UV slopes of LBGs at similar redshifts and luminosities have been presented by \cite{bou2009}, \cite{fink2012}, and \cite{hathi2013}, albeit on samples smaller than ours and using slightly different fitting techniques. We do not consider here the analysis by \citet{Kurczynski2014} and \citet{Pannella2015}, which are mostly focused on fainter galaxies.

\cite{bou2009} also suggested a Gaussian distribution and found a mean value $<\beta>\simeq-1.18$ for a colour-selected sample of 168 bright LBGs ($<M_{UV}>=-21.73$) at redshift $z\simeq 2.5$. Their mean UV slope is redder than our estimate, probably as a result of the lower redshift range probed and because they adopted different colour-selection criteria.

\cite{fink2012} exploited a sample of 177 galaxies selected on the basis of their photometric redshift and with -22 $\lesssim M_{1500}\lesssim -16$. Their sample has a mean photometric redshift of $3.42$. 
They measured UV slopes on the best-fit templates of their objects and find a median $\beta$ of $-1.82_{-0.04}^{+0.00}$ that shifts to $-1.80_{-0.06}^{+0.03}$ when the brightest galaxies of the sample are considered ($L>0.75L^{*}$). These results are in agreement within the uncertainties with our findings in the same magnitude range (second magnitude bin in Tab.~\ref{tab2}). Our redder mean value can be explained by the small difference in redshift between the two samples, or by the different methodologies adopted.

\cite{hathi2013} analysed a sample of LBGs at $z\simeq 2.6$ and $z\simeq3.8,$ finding $<\beta>=$-1.71 and $<\beta>=$-1.88, respectively. They obtained these results by fitting a Gaussian to the observed $\beta$ distribution to find the median value. At variance with our procedure, they did not take into full account the biases that are introduced by observational and selection effects. When we applied their simpler approach to our data, we find a mean $<\beta>\sim -1.82$. This value falls in the middle of their estimates, in agreement with the well-established trend of decreasing typical $\beta$ at increasing redshift.

\begin{figure}
 \centering
 	\includegraphics[width=8.0cm]{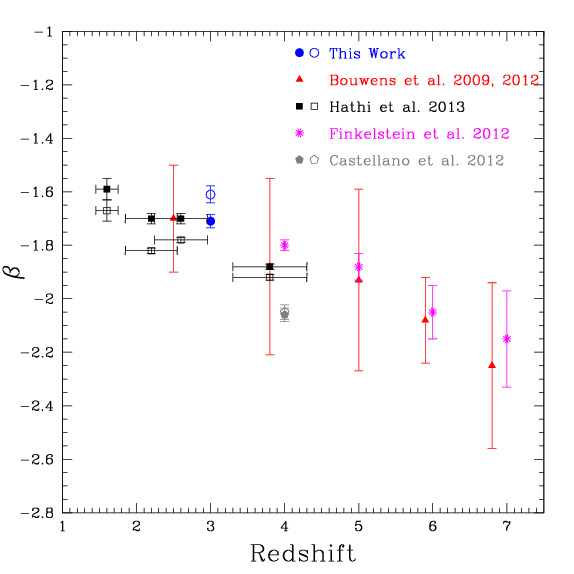}
        \caption{Mean UV slope as a function of redshifts for $ L\gtrsim L^*$ galaxies from this work and from the literature. Estimates from \cite{fink2012} are for the $L>0.75 L^*$ range (Tab. 4 in their paper), while points from \cite{bou2012} are for objects with average $M_{UV}=-20.5$ up to z=6, and $M_{UV}=-20.25$ at z=7 (Tab. 4 in their paper).   Filled and open symbols for this work and \cite{cast2012} indicate the Gaussian and the log-normal best-fit values, respectively.  The median $\beta$ values from \cite{hathi2013} based on UV slope fits in the rest-frame 1300 - 1900\ \angstrom\ and 1300 - 3400\ \angstrom\  wavelength ranges are indicated as filled and open squares, respectively. Error bars for all the points indicate the $1 \sigma$ uncertainty on the mean, except for the \cite{bou2009} and \cite{bou2012} z=4-5 values, for which they are representative of the $1 \sigma$ scatter. }
        \label{fig13}
 \end{figure}

We first compare in Fig.~\ref{fig13} our results to the estimates at z$\sim$3 and to estimates at lower and higher redshifts in the same luminosity range from \cite{bou2012}, \cite{fink2012} and C12.
Our results properly fit in the well-established trend of decreasing UV slopes at increasing redshift and at fixed UV luminosity. In particular, a self-consistent comparison can be made with the results by C12 at $z\sim 4,$ where the very same method for recovering the intrinsic UV slope distributions was applied. The brightest galaxies in the C12 sample ($M_{UV}\lesssim -21$) have $<\beta>=-1.90$, implying that in the $\sim 500 Myr$ elapsed between redshift 4 and 3.2 the typical UV slopes increase by $\Delta \beta \simeq 0.2$, corresponding to an increase in extinction $A_{1600}$ of $\simeq$0.4 magnitudes assuming the insights of \citet{meu1999}. Interestingly, there is a significant evolution in the scatter of the distribution that increases from $rms(<\beta>)=0.35$ at z$\sim4$ to $rms(<\beta>)=0.56$ found in our sample. While an increase in dust extinction is the most likely explanation for this evolution, only a thorough investigation including a deep spectroscopic analysis \citep[e.g.][]{Castellano2014} can distinguish the contribution of dust and other factors (in particular, age and metallicity) to the observed trends. Moreover, this evolution between $z\sim3$ and $z\sim4$ seems to be stronger than the evolution suggested by measurements at other redshifts shown in Fig. \ref{fig13}.

The interpretation of the observed redshift-$<\beta>$ trend at fixed luminosity is not straightforward because it may result from an evolution with redshift of the typical stellar mass of $L\sim L^*$ galaxies, given an underlying relation between stellar mass and dust extinction \citep{Pannella2009,Reddy2010,Pannella2015,Whitaker2014,Bouwens2016}. We therefore compared our sample to similar samples that are available in the literature after estimating stellar masses of our objects through a spectral energy distribution (SED) fitting approach. We used the \verb|zphot.exe| code and \citet{Bruzual2003} templates following the technique described in \citet{Castellano2014} and \citet{Castellano2016}, finding masses in the range $log(M_{star}/M_{\odot})=9 - 11$ and an average $log(M_{star}/M_{\odot})=9.8 \pm 0.6$ \citep[][stellar IMF]{Salpeter1955} for the galaxies in our sample. Our $<\beta> \simeq -1.7$ is slightly redder than the extrapolation of the $\beta$-stellar mass at z=4 presented by \cite{fink2012}: they find an increasing slope at increasing stellar mass, with a median $\beta=-1.88$ at z=4 and a nearly constant $\beta \sim -1.8 $ at z=5-7 for objects with $log(M_{star}/M_{\odot})=9 - 10$. At z$\sim$2, \citet{reddy2018} found slopes that are redder than in our z$\sim$3 sample: they measured $<\beta>$ = -0.92 (-1.88) for objects with masses higher (lower) than $log(M_{star}/M_{\odot})=9.75$. We verified that our average $<\beta> \sim -1.7$  does not significantly change when the sample is restricted to only the sources with $log(M_{star}/M_{\odot})>9.75$. Overall, these results suggest that the relation between stellar mass and UV slope may experience a mild evolution up to z$\sim$3 and a stronger one from z$\sim$3 to $\sim$2. This trend is reminiscent of theoretical predictions on the mass-metallicity relation where a very mild evolution is found at z$\gtrsim$3-4 \citep[][]{Mitra2015,Dave2017}. Assuming that a fixed fraction of metals are incorporated in dust, as suggested by observations at both low and high redshift \citep[e.g.][]{Draine2007,Chen2013}, little evolution of the relation between mass and UV slope would be related to a mild increase at high redshift of the metallicity at fixed stellar mass.

\begin{table*}       
\centering      
\begin{adjustbox}{max width=\textwidth}             
\begin{tabular}{cccc} 
\hline    
\hline                 
log(SFRD) & Fraction &$A_{1600}-\beta$ conversion& Extinction law\\   
 $M_{\odot} / yr / Mpc^{3}$& & &\\ 
\hline              
$-1.63_{-0.10}^{+0.21}$& 25\% & \cite{meu1999}& Starburst$^a$\\
$-1.28_{-0.10}^{+0.11}$& 59\%& \citet{Castellano2014} &  Starburst$^a$\\
$-1.35_{-0.12}^{+0.10}$& 50\%& \citet{deBarros2014} &  Starburst$^a$\\
$-1.33_{-0.11}^{+0.10}$& 52\%& \citet{Cullen2017} &  Modified Starburst$^b$\\
$-1.53_{-0.13}^{+0.12}$& 32\%&  \cite{gall2010} & MEC$^c$ \\
$-1.96_{-0.05}^{+0.04}$& 12\%&  \cite{bouw2016} & SMC$^d$\\
$-1.82_{-0.04}^{+0.05}$& 17\%&  \cite{reddy2018} & SMC$^d$\\
\hline      
\end{tabular}
\end{adjustbox}
\caption{Contribution from $L>L^*$ galaxies to the SFRD at z$\sim$3 under different assumptions. A Gaussian UV slope distribution and Schechter LF parameters from \citet{Reddy2009} are assumed in all cases. The reference total SFRD is taken from \citet{MadDick2014}. a) \cite{calz2000}, b) \citet{Cullen2017}, c) \cite{gall2010}, and d) \cite{Gordon2003}.}
\label{tabSFRD}
\end{table*} 

\section{Dust extinction and corrected SFR in bright z$\sim$3 LBGs}
\label{section5}

\subsection{Star-formation rate density of $L>L^*$ galaxies}
The physical interpretation of our results is not straightforward because many factors affect the UV slope, although extinction is by far the main influencing factor (e.g. \citealp{wilk2011}). We can convert our $\beta$ values into the average dust extinction at $1600\angstrom$ assuming standard relations. Following the widely adopted \cite*{meu1999} relation, we find $A_{1600}\simeq 1.07 $ ($A_{1600}\simeq 1.03$) and $A_{1600}\simeq 1.01$ ($A_{1600}\simeq 1.45$) for the Gaussian (log-normal) distribution in the bright and in the faint bin, respectively.

We can then evaluate the contribution from ultra-bright LBGs to the SFRD on the basis of the estimated PDF($\beta$) through the following equation:
\begin{equation}
SFRD=\frac{1.0}{8\times 10^{27}}\int dL \int dA\cdot PDF(A,L) 10^{0.4\cdot A}\cdot L\cdot\Phi(L)
,\end{equation}
where the constant factor is from \cite{mad1998}, $A = A_{1600}$ such that $PDF(A, L)$ is univocally related to the $PDF(\beta, L)$ through the \cite{meu1999} relation, and $\Phi(L)$ is the UV luminosity function at $1600 \angstrom$ \citep[we adopt Schechter parameters from][]{Reddy2009}.  The integral is evaluated separately for each of the two magnitude bins considering the relevant best-fit distribution $PDF(\beta)$, and the sum of the resulting values yields the SFRD at $L>L^*$ at z=3. We obtain $log(SFRD) = -1.63_{-0.10}^{+0.21} M_{\odot} / yr / Mpc^{3} $ for the Gaussian $PDF(\beta)$ and $log(SFRD)=-1.57_{-0.06}^{+0.06} M_{\odot} / yr / Mpc^{3}$  for the log-normal PDF at $M_{UV}\leq -21.0$, which is the magnitude range probed by our data. 
 
On the basis of the best-fitting SFRD function from \cite{MadDick2014} (Fig. 9 and Eq. 15), $L>L^*$ LBGs contribute $\sim 25\%$ of the global value. This estimate is of course an upper limit if a non-negligible contribution to the SFRD at these redshifts came from optically undetected sub-millimeter galaxies (SMGs), such as those found by \cite{sant2016} at $z\sim 3.3$ and \cite{dad2009} at $z\sim 4$.

Unfortunately, the $A_{1600}$ and SFRD estimates strongly depend on assumptions on the attenuation curve and on the $A_{1600}-\beta$ relation. The latter depends on the intrinsic UV slope of the source, which in turn is determined by metallicity, age, SFH, and IMF of its stellar populations \citep[e.g.][]{Castellano2014,deBarros2014}. We investigated how uncertainties in dust attenuation properties affect our SFRD estimate by adopting different attenuation curves and $A_{1600}-\beta$ conversion laws from the literature when we converted our $PDF(\beta, L)$ to the $PDF(A, L)$ used in Eq. 1.   When we adopted the $A_{1600}-\beta$ relation by \citet{Castellano2014} (Eq. 1 in their paper), which takes into account the low stellar metallicity measured in bright z$\sim$3 LBGs (intrinsic UV  slope $\beta_{dust-free}$=-2.67), we find a higher extinction ($A_{1600}\simeq 1.93$ for the Gaussian distribution) resulting in $  \text{an approximately twice higher}$  SFRD ($log(SFRD) = -1.28_{-0.10}^{+0.11} M_{\odot} / yr / Mpc^{3}$). The $A_{1600}-\beta$ relation from \citet{deBarros2014} (intrinsic UV  slope $\beta_{dust-free}$=-2.58) yields $log(SFRD) \simeq -1.35 M_{\odot} / yr / Mpc^{3}$, which amounts to a 50\% contribution to the global SFRD. A similar result is found with the best-fitting dust model estimated at z$\sim$5 by \citet{Cullen2017}.
Similarly, SFRD estimates ~25\% higher than our reference value are found when the \cite{gall2010} attenuation law is assumed, which is flatter than that of \cite{calz2000}, while the assumption of a Small Magellanic Cloud extinction leads to values ~53\%  or ~35\% lower when the $A_{1600}-\beta$ conversions from \citet{bouw2016} and \citet{reddy2018} are adopted, respectively. The results are summarised in Table~\ref{tabSFRD}. Uncertainties in the LF parameters also yield a comparable uncertainty on the SFRD: our estimates have to be revised upward by $\sim$30\% when the Schechter parameters from \citet{Cucciati2012} are adopted, while a $\sim$15\% increase is found when the LF from \citet{Mehta2017} or that from \citet{Viironen2018} is used.

\subsection{UV slope as a function of dust-corrected UV luminosity}
\label{section6}
\begin{figure}
\centering
	\includegraphics[width=8.0cm]{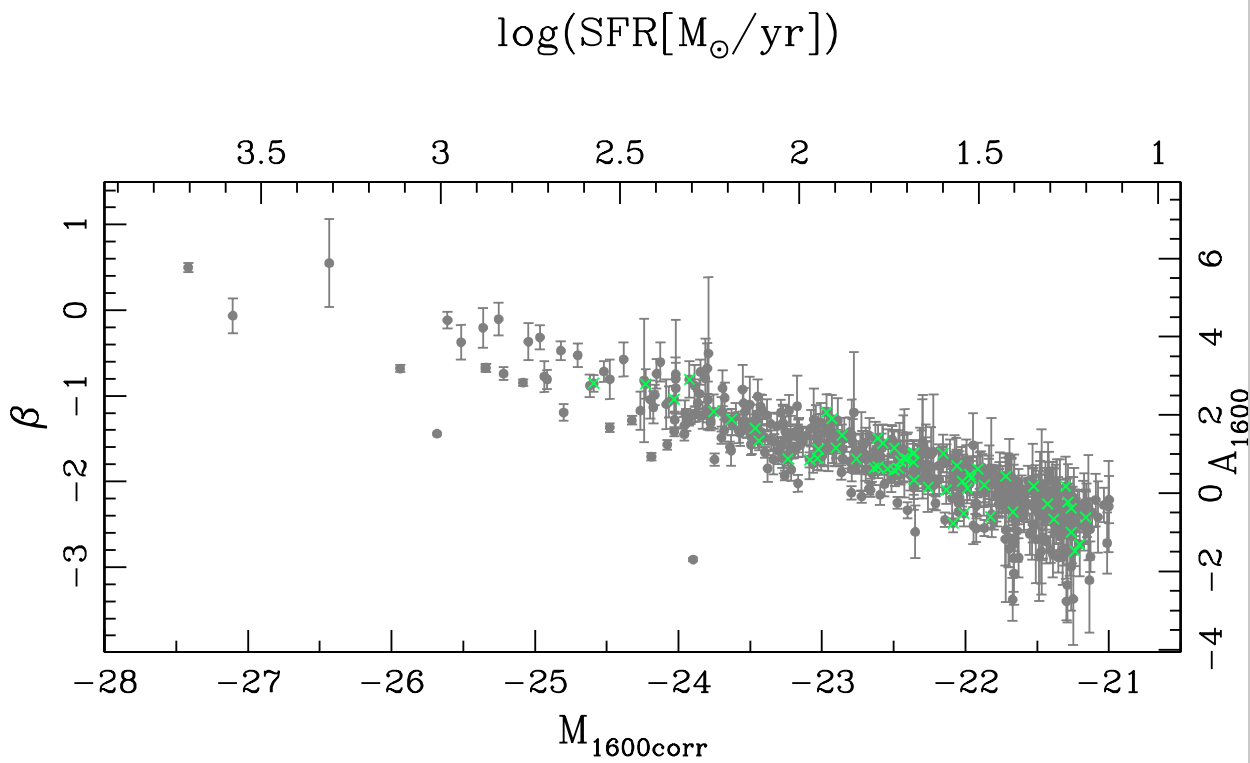}
        \caption{UV slope as a function of dust-corrected UV magnitude. The figure also shows the conversion of these two quantities into extinction $A_{1600}$ \citep{meu1999} and $log(SFR)$ \citep{mad1998}. Green crosses mark objects for which a spectroscopic redshift is available.}
        \label{fig14}
\end{figure}

The usual analysis of trends between $\beta$ and $M_{UV}$ is mainly a useful tool to evaluate the correction to the observed UV luminosity functions and, eventually, to the SFRD \citep[e.g.][and references therein]{Dunlop2013review,MadDick2014}. However, it does not provide a clear mean to constrain the physical properties of the LBGs because dust affects both the UV slope and the observed UV magnitude. 
In order to evaluate the effect of dust as a function of an intrinsic galaxy property, we estimated the relation between $\beta$ and the UV rest-frame magnitude corrected for the extinction $M_{1600corr}=M_{1600}-A_{1600}$. Figure \ref{fig14} shows this relation along with a conversion of $\beta$ and $M_{1600corr}$ into $A_{1600}$ and SFR, respectively, estimated following \citet{meu1999} and \citet{mad1998} (the adoption of different relations implies a rescaling of the units). Of course this conversion is still affected by a degeneracy, being $A_{1600}$ estimated from $\beta$; nevertheless it allows a direct visualization of how dust extinction evolves for galaxies at different luminosities. 
A clear trend seems to be in place, with extinction decreasing at increasing $M_{1600corr}$.
The brightest galaxies are the dustiest and most star-forming, with $A_{1600}\gtrsim 2$, and $SFR\gtrsim300 M_{\odot}/yr$ at $M_{1600corr}\lesssim -24.5$. Our findings are in agreement with the analysis of the $M_{1600corr}-\beta$ relation at $z\sim 4$ presented in C12, and are consistent with a scenario where the more massive galaxies are more dust attenuated \citep[e.g.][]{Pannella2015,bouw2016}.

\section{Summary and conclusions}
\label{section7}
We have presented the analysis of the UV slopes of a large sample of bright $z\sim3$ LBGs in the COSMOS field. We produced a photometric catalogue spanning the wavelength range from $335nm$ to $2150nm$ and exploited a new technique alternative to PSF-matching that maximises the S/N of colour measurements.

A selection criterion based on the U, G, and Y bands was used to select 517 $z\sim 3$ R-detected objects at $S/N(R)\geq20$. The deep Y band enables a selection that is more efficient than can be achieved with the standard UGR criterion: in particular, a check on spectroscopically confirmed sources shows that our selection is characterized by a lower contamination from lower redshift interlopers and recovers more objects at z$\simeq$2.7-3.3. We measured the $\beta$ slopes by performing a linear fit on the I, Z, Y magnitudes  and evaluated observational uncertainties using extended simulations.

We carried out a parametric analysis of the UV slope distribution under the assumption that the intrinsic probability distribution function PDF($\beta$)  is either Gaussian or log-normal. The best-fit values were found by comparing the observed and the simulated counts through a maximum  likelihood estimator. We find a typical average $<\beta>\simeq -1.70$, with a dispersion rms($\beta$)$\simeq$0.55. 
When comparing with the values measured at z$\sim$4 with the same technique, we note a significant increase both in the average ($<\beta>\simeq -1.90$ at z$\sim$4) and in the scatter of the distribution (rms($\beta$)$\simeq$0.35 at z$\sim$4). Our results fits well in the relation of $\beta$ versus redshift, where galaxies are found to be bluer at increasing redshifts. The resulting contribution of ultra-bright LBGs to the z$\sim$3 SFRD ($log(SFRD)=-1.63_{-0.10}^{+0.21}M_{\odot}/yr/Mpc^{3}$ for the Gaussian PDF($\beta$)) corresponds to $\sim25\%$ of the global value. We also found a clear trend with decreasing extinction at increasing intrinsic UV magnitude.

\begin{acknowledgements}
We thank the anonymous referee for the useful suggestions and constructive comments that helped us to improve this paper. The research leading to these results has received funding
from the European Union Seventh Framework Programme (FP7/2007-2013) under grant agreement No. 312725. The present work is based on observations carried out at
the Large Binocular Telescope at Mt. Graham, AZ. The LBT is an international collaboration among institutions in the United
States, Italy, and Germany. LBT Corporation partners are The University of Arizona on behalf of the Arizona university
system; Istituto Nazionale di Astrofisica, Italy; LBT Beteiligungsgesellschaft, Germany, representing the Max-Planck
Society, the Astrophysical Institute Potsdam, and Heidelberg University; The Ohio State University; and The Research
Corporation, on behalf of The University of Notre Dame, University of Minnesota, and University of Virginia.
\end{acknowledgements}
\appendix

\section{Photometric catalogue}
\label{section8}

\subsection{New photometric method for estimating unbiased colours from images with different seeing}

The images in our dataset have been acquired using different telescopes under a variety of observing conditions, which naturally leads to different point spread functions (PDFs). In this situation, the shape of a given object is smoothed in different ways in the various images, with the result that a standard fixed aperture photometry yields severely biased colours because apertures of the same size recover a different flux fraction in each image. The standard approach to deal with these inhomogeneous datasets is to match the PSFs through appropriate kernels (e.g. \citealp{cast2010}, \citealp{gal2013}, \citealp{guo2013}. \citealp{Mer2016}). 

We present here a different method that does not imply the usual PSF-matching and thus avoids the loss of information arising from the degradation of higher resolution images to match lower resolution images. Our technique aims at defining apertures that enable a self-consistent flux measurement on images with different PSFs. It is of clear advantage in the analysis of UV slopes to achieve unbiased colour estimates at higher S/N than thos of standard PSF-matched catalogues.\textup{\textup{\textup{}}}

The first step is the definition for each image of an \textit{effective FWHM} ($FWHM_{eff}$) enclosing a fixed fraction of the total flux of bright non-saturated stars, regardless of the exact shape of the PSF:

\begin{equation}
\label{eq1}
\frac{\texttt{FLUX\_AUTO}}{\texttt{FLUX\_APER}}=1.5
\end{equation}
where \texttt{FLUX\_AUTO} is the Kron flux of the source \citep{bertarn1996} and \texttt{FLUX\_APER} is the flux measured in a circular aperture with a diameter equal to $FWHM_{eff}$. The ratio was chosen as a trade-off between having enough flux at a reasonable S/N from very faint objects as well and avoiding too  large and noisy apertures.  

In practice, the $FWHM_{eff}$ value was found by successive approximations on each image of the dataset, and it is reported in Table \ref{tabA1}. The effective FWHM (which is different for each band) provides a reference unit that characterises the PSF in a given image. 

\begin{table}  
        \centering      
        \begin{adjustbox}{max width=\textwidth}             
                \begin{tabular}{cc} 
                        \hline    
                        \hline                 
                        Filter & $FWHM_{eff} (arcsec)$   \\    
                        \hline                        
                         $U_{special}$ & 1.37    \\     
                         $G_{sloan}$ & 1.80    \\
                         $R_{sloan}$ & 1.43    \\ 
                         I$_{sloan}$ & 1.28    \\ 
                         $Z_{sloan}$ & 1.59    \\ 
                         $Y$ & 1.67    \\ 
                         $J$ & 1.48    \\
                     $H$ & 1.34   \\
                         $K$ & 1.13    \\
                        \hline      
                \end{tabular}
        \end{adjustbox}
        \caption{$FWHM_{eff}$ of the images.}
        \label{tabA1}
\end{table}

\begin{figure}
\centering
	\includegraphics[width=8.0cm]{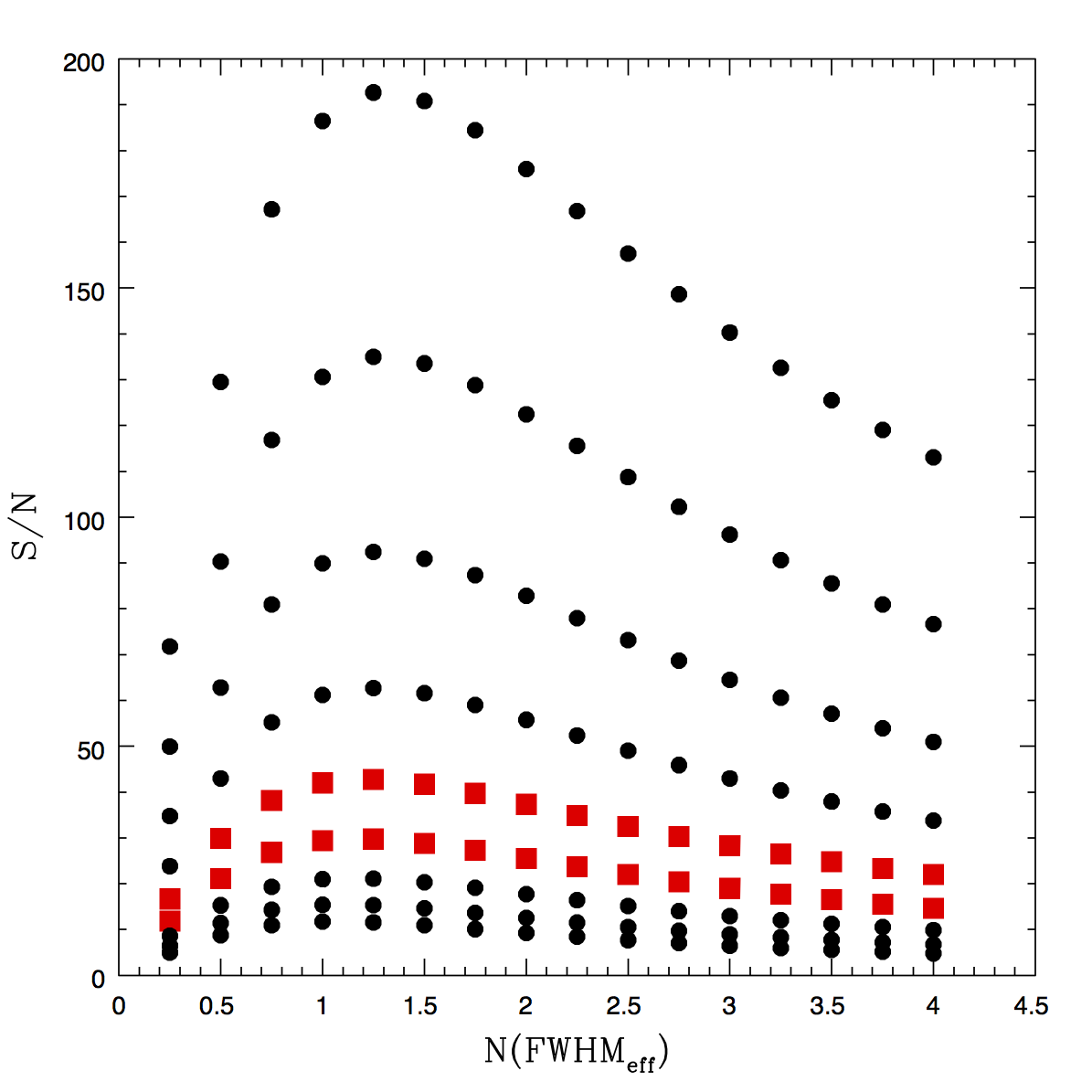}
        \caption{Mean R-band S/N in different photometric apertures defined in terms of the $FWHM_{eff}$, for objects in nine different magnitude bins at $22.0 \lesssim R\lesssim 26.5$ (from top to bottom). The $24.0 \lesssim R\lesssim 24.5$ and the $24.5 \lesssim R\lesssim 25.0$ bins where the bulk of our sources are found are shown as red filled squares.}
        \label{fig1}
\end{figure}

The second step was to identify the optimum aperture in the detection image (R band in our case) that is designed to maximise the S/N of the photometric measurement for the objects of interest. Figure \ref{fig1} shows the S/N as a function of aperture diameter (in terms of times the $FWHM_{eff}$) for sources at different magnitudes (ranging from 22 to 26.5 mag at 0.5 mag intervals): the aperture maximising the S/N extends from 1$\times$ to $1.5\times FWHM_{eff}$ at decreasing flux. Because the bulk of the sources in our sample has magnitude in the range $24.0 \lesssim R\leq 24.8,$ we set the optimum aperture in this band to $1.25\times FWHM_{eff}$. 

The third and final step was the identification of matched apertures in the measurement images, that is, the apertures (in units of the relevant $FWHM_{eff}$) that enable measuring unbiased colours on all sources of interest.
This was done through intensive image simulations based on the $H_{160}$ image from the GOODS-ERS WFC3/IR dataset (HST Programme ID 11359, \citealp{wind2011}), which has a uniform depth and PSF ($FWHM_{H_{160}}=0.18\ arcsec$) across $\sim$45 arcmin$^2$. By smoothing this image with appropriate convolution kernels, we obtained from it images with a seeing that ranged from $0.6\ arcsec$ to $1.2\ arcsec$ at $0.1\ arcsec$ intervals. Clearly, all objects are expected to have a colour term equal to zero inall images in this dataset.

We then ran SExtractor on dual-image mode on all images and searched for the apertures that yielded the expected colours. 
We found the following empirical relation between the size of the optimal aperture ($Aper_{meas}$) and the $FWHM_{eff}$ of the detection and measurement images:

\begin{equation}
\label{eq2}
Aper_{meas}=0.4 \times [FWHM_{eff}(det)-FWHM_{eff}(meas)]+1.25
.\end{equation}

\begin{figure}
	\centering
	\includegraphics[width=8cm]{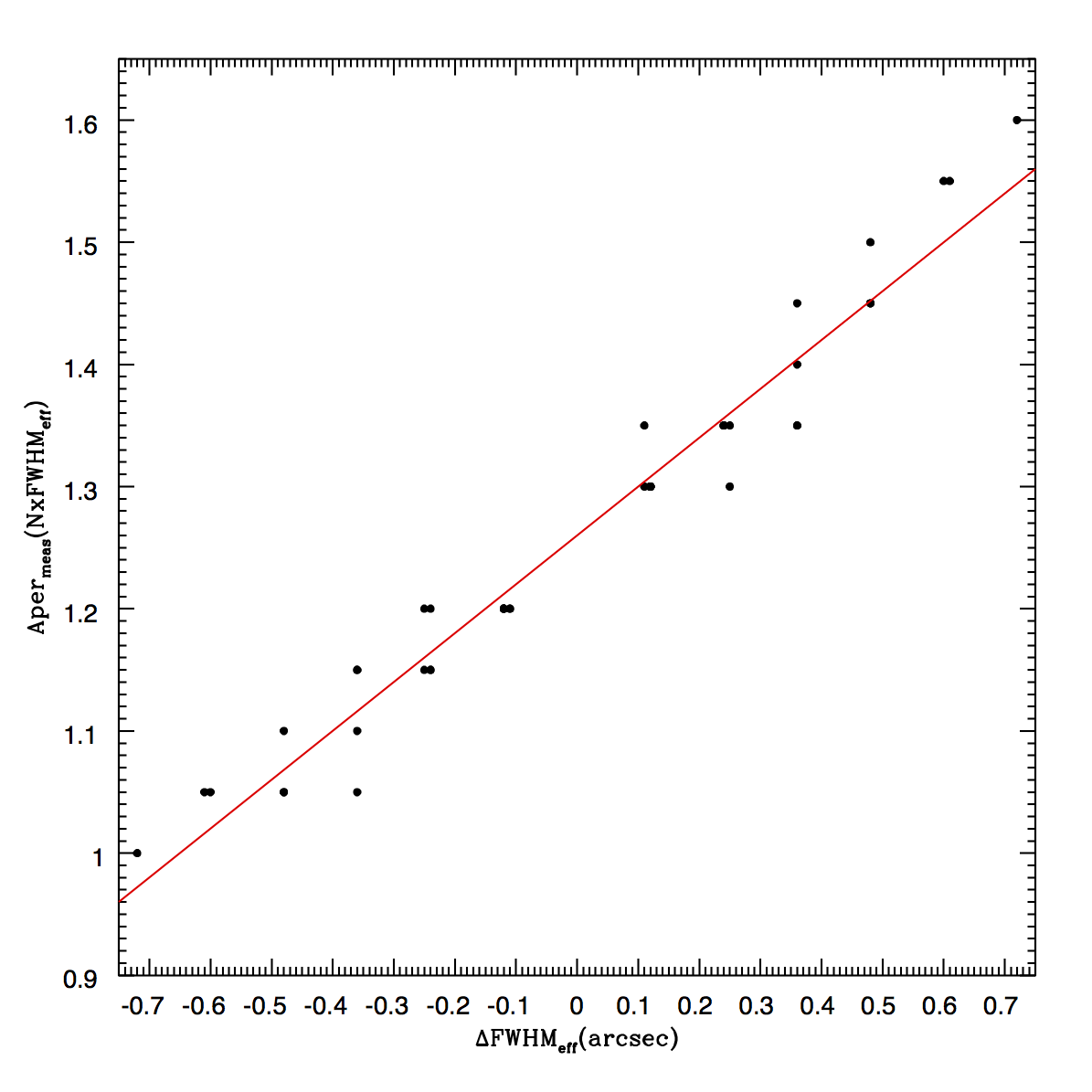}
        \caption{Empirical relation \ref{eq2}. Black dots show the size of the photometric aperture in units of $FWHM_{eff}$ of the measurement image. The red line shows the linear fit to the data.}
        \label{fig3}
\end{figure}

Fig.~\ref{fig3} shows the relation in Eq.\ref{eq2}: the red line represents a linear fit to the data with slope and intercept equal to $0.4$ and $1.26,$ respectively. 

\subsection{Multi-band catalogue for the COSMOS dataset}
The procedure described above was used in the analysis of our COSMOS dataset. We used the R-band mosaic as detection image and ran SExtractor in dual-image mode to measure the magnitudes in all bands. The magnitude of the sources in each band was set to

\begin{equation}
\label{eq3}
mag_{filter}=R_{TOT}+(mag_{aper,filter}-mag_{aper,R}),
\end{equation}
where total magnitudes in the R band were computed using SExtractor \texttt{MAG\_AUTO}, and apertures were defined following Eqs.~\ref{eq1} and \ref{eq2}. 

\begin{figure}
\centering
	\includegraphics[width=8.0cm]{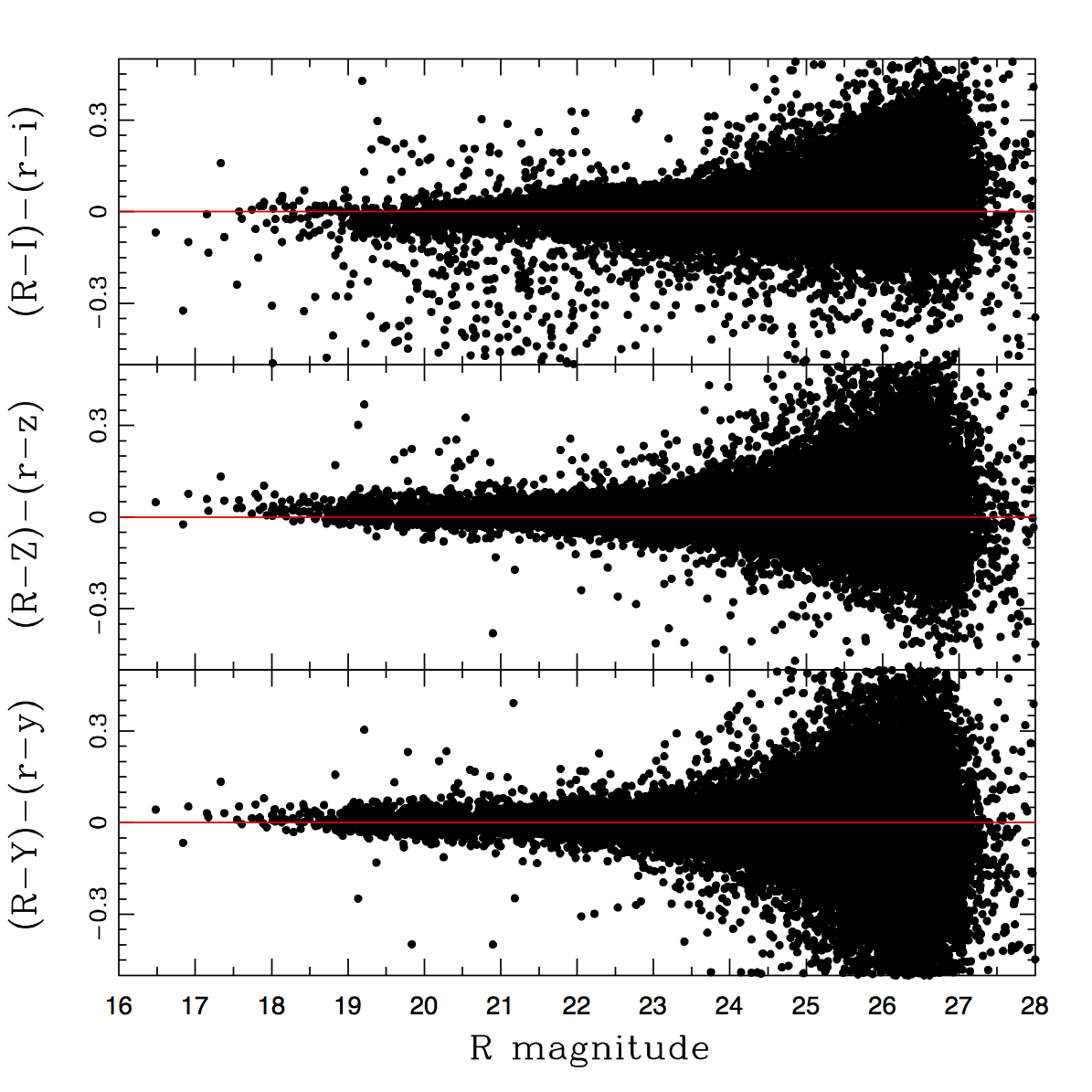}
        \caption{Comparison between colours from our catalogue and those from a catalogue built using the standard PSF-matching procedure. Capital letters are used for the magnitudes in our catalogue, and lower case letters indicate the magnitudes in the PSF-matched catalogue. The bulk of bright outliers at R$\sim$19-23 in the top panel are found to be saturated stars in the I-band image.}
        \label{fig4}
\end{figure}

We tested our procedure by comparing the resulting catalogue to an R-detected catalogue extracted following the standard \textup{PSF-matching} approach. In the latter case, total magnitudes are extracted following equation \ref{eq3}, where the colour term was instead evaluated between images smoothed to the Z-band PSF, which is the coarser in the dataset. We checked the accuracy of the PSF-matching procedure by measuring the PSF growth curve for each smoothed image. We found that the matching is accurate to within $3\%$ at  $\text{twice the }$FWHM, which is the fixed size within which we measure colours in this case.  Fig.\ref{fig4} shows a comparison between the colours evaluated with the two different methods. Only colours between bands that are involved in the measurement of $\beta$ are shown; all other combinations were tested as well and show similar results.

\begin{figure}
\centering
	\includegraphics[width=8.0cm]{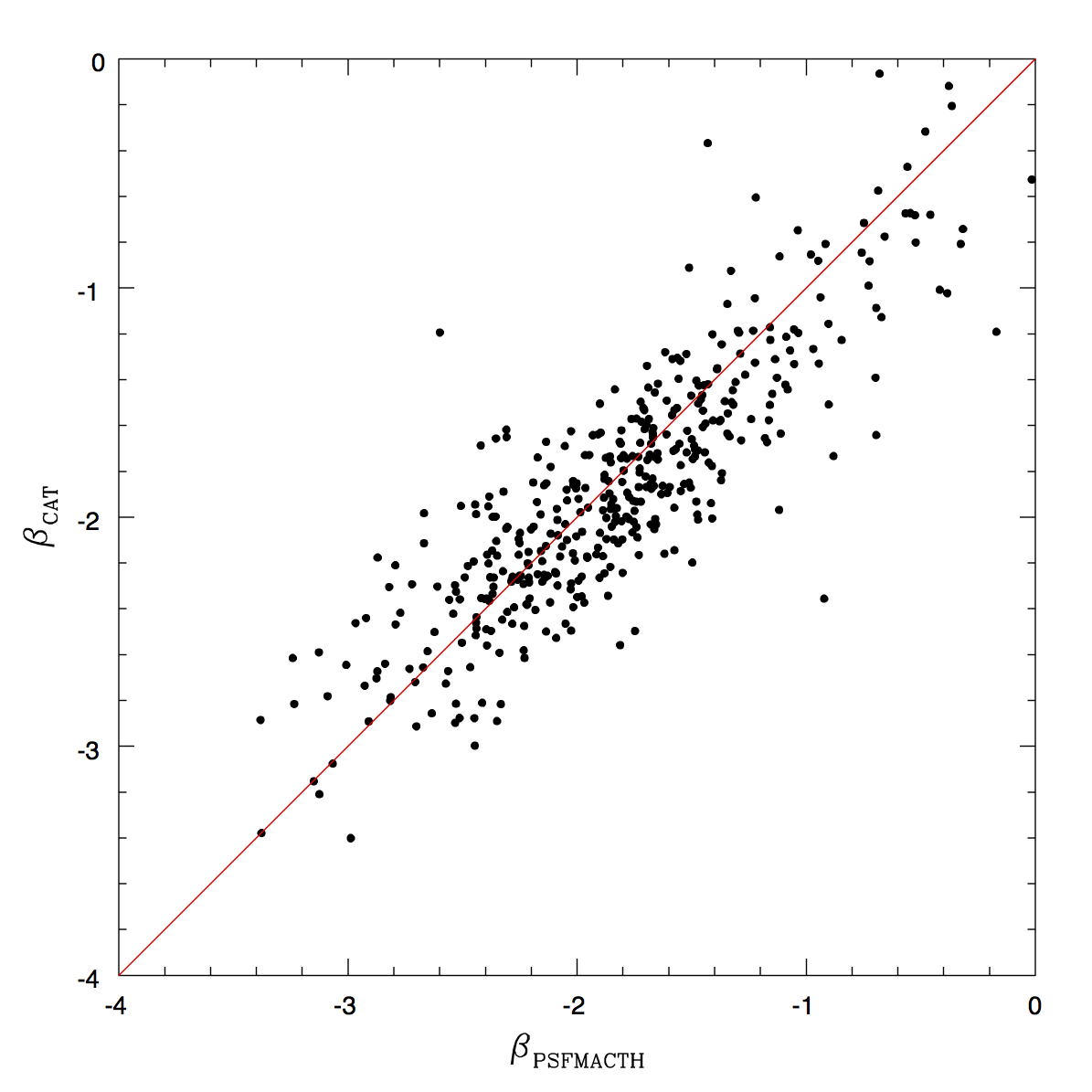}
        \caption{Comparison between UV slopes from our catalogue and those from a catalogue built using the standard PSF-matching procedure. The red line marks the equality relation.}
        \label{fig22}
\end{figure}

In order to quantify possible biases introduced by our technique in the measurement of $\beta$, we re-estimated UV slopes for all objects in our sample using magnitudes from the PSF-matched catalogue. We find an average  $<\beta_{PSFMATCH}>=-1.81 \pm 0.03$, perfectly consistent with the $<\beta>=-1.82 \pm 0.03$ found in our reference catalogue.
In Figure \ref{fig22} we show a comparison between the two UV slope estimates for all objects. 

All these tests show no evident systematics between our photometric method and the standard PSF-matching procedure. Small differences in photometry on single sources are most likely due to
\begin{itemize}
        \item{imperfections in the PSF extraction or in the creation of the PSF-matching kernels}
        \item{the uncertainty in the evaluation of $FWHM_{eff}$ on which our photometric technique is based.}
\end{itemize}
These issues are under investigation and will be further improved. In particular, our technique can be improved by adopting elliptical apertures that will allow us to better follow the galaxy profiles, and by an automatic search of  apertures that maximise the S/N \citep[e.g. APHOT,][]{Merlin2019}. 

\begin{figure}
\centering
	\includegraphics[width=8.0cm]{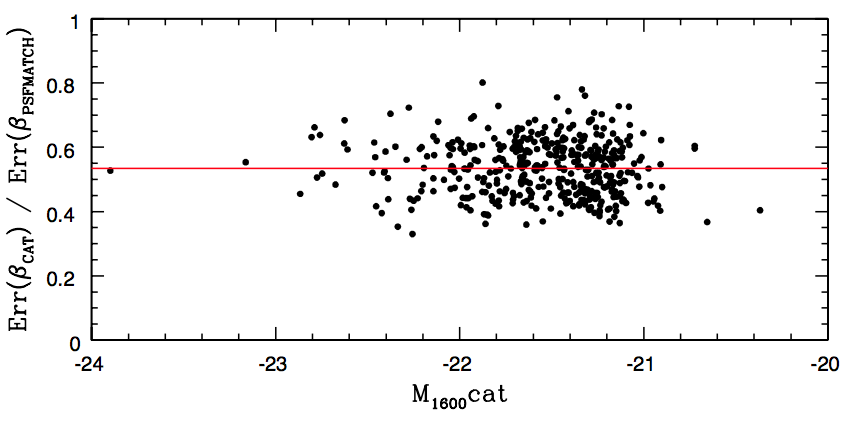}
        \caption{Ratio between the uncertainty on UV slope measured from our reference catalogue and from the PSF-matched catalogue as a function of UV rest-frame magnitude of the sources. The red line at $~0.53$ marks the average value of the ratio.}
        \label{fig24}
\end{figure}

Finally, we compared the uncertainty on $\beta$ between the two different approaches object by object. Figure \ref{fig24} shows that errors on the UV slopes measured from our catalogue are lower by a factor of $\sim 2$ (on average) than in the case when the PSF-matched catalogue is used.  This shows that our approach enables measuring \textit{\ss}\ at a higher S/N than when previous photometric procedures are adopted.

\bibliographystyle{aa}

\end{document}